\documentclass[rnote]{aa} % for the research notes
\usepackage{graphicx}
\usepackage{natbib}
\usepackage[varg]{txfonts}
\begin{document}
\titlerunning{Atom-Rydberg atom collisions in
helium-rich white dwarf atmospheres}
\authorrunning{Sre{\'c}kovi{\' c} et al.}
   \title{Excitation and deexcitation processes in atom-Rydberg atom collisions in
helium-rich white dwarf atmospheres}

   %\subtitle{}

   \author{V. A. Sre{\'c}kovi{\' c}
          \inst{1}
          \and
            A. A. Mihajlov
          \inst{1}
          \and
             Lj.M. Ignjatovi{\' c}
           \inst{1}
          \and
          M. S. Dimitrijevi{\' c}\inst{2}
          }

   \institute{University of Belgrade, Institute of physics, P.O. Box 57, 11001, Belgrade, Serbia;
\\and
       Institute Isaac Newton of Chile, Yugoslavia Branch, Belgrade, Serbia\\
              \email{vlada@ipb.ac.rs}
         \and
             Astronomical Observatory, Volgina 7, 11060 Belgrade 74
       Serbia;\\
       and Institute Isaac Newton of Chile, Yugoslavia Branch, Belgrade, Serbia;\\ and
       Observatoire de Paris, 92195 Meudon Cedex, France\\
             %\thanks{The university of heaven temporarily does not
%                     accept e-mails}
             }

   \date{Received ; accepted , }

  \abstract
   {We aim to show the importance of non-elastic excitation and
deexcitation processes in $\textrm{He}^{*}(n)+ \textrm{He}(1s^{2})$ collisions with the principal quantum number
$n \ge 3$ for helium-rich white dwarf atmospheres. We compare the
efficiencies of these processes with those of the known non-elastic electron-$\textrm{He}^{*}(n)$ atom
processes in the atmospheres of some DB white dwarfs. We show that in significant parts of the considered
atmospheres, which contain weakly ionized layers (the ionization degree $\lesssim 10^{-3}$), the
influence of the studied atom-Rydberg atom processes on excited helium atom populations is
dominant or at least comparable to the influence of the concurrent electron-$\textrm{He}^{*}(n)$-atom
processes.}


   \maketitle
%
%________________________________________________________________

\section{Introduction}

We here continue our previous investigations of inelastic
$A+A^{*}(n)$ atom-Rydberg atom collision processes (where the principal quantum
number was $n>>1$ and $A$ was the ground-state atom), which influence the
excited atom populations in stellar atmospheres. We consider the processes that can
be interpreted as resulting from a resonant energy exchange between the subsystem
$A+A^{+}$, in which $A^{+}$ is the core of the Rydberg atom $A^{*}(n)$, and the
outer electron of this atom. There are two groups of such
processes. The first group involves the chemi-ionization processes \citep{jan80,mih96,mih97},
which connect the block of atomic Rydberg states with the continuum. The
second group comprises the so-called $(n-n')$-mixing processes within
the above mentioned block of atomic Rydberg states \citep{mih82,mih04,mih08}, i.e. the excitation and
deexcitation processes that imply transitions between the Rydberg states with
the principal quantum numbers $n$ and $n' \ne n$.

The chemi-ionization processes in $A^{*}(n)+A$ collisions, together with their
inverse processes, have been studied for both $A=\textrm{H}$ and $A=\textrm{He}$ cases, which
are especially important for stellar atmospheres. \citet{mih03, mih07}
studied the influence of these processes on the excited hydrogen atom
populations and the spectral line profiles in the atmosphere of an M red dwarf
($T_{eff}=3800$K) using the PHOENIX code. \citet{mih97,mih11} and \citet{mih03} studied the potential
influence of these processes on the excited hydrogen and helium atom
populations in the atmospheres of Sun and of some DB white dwarfs.

However, the possible influence of the $(n-n')$-mixing processes in
$A^{*}(n)+A$ collisions on the excited-atom populations in stellar atmospheres
has only been studied for the $A=\textrm{H}$ case in the context of the atmosphere of
the Sun \citep{mih05}. In the above mentioned studies concerning the solar
atmosphere, the necessary calculations were performed within the
corresponding non-local thermodynamic equilibrium (non-LTE) model proposed by \citet{ver81}.
The findings of these studies have shown that the efficiency of these processes in a major part of the
solar photosphere is dominant with respect to the efficiency of the relevant
concurrent excitation and deexcitation processes for $n=4,5$ and $6$, and is
comparable with them for $n=7$ and $8$. The above mentioned studies concluded that
the excitation and deexcitation processes in $\textrm{H}^{*}(n)+\textrm{H}$ collisions should be
included \emph{ab initio} into future non-LTE models of the solar
atmosphere. The importance and effects of the atom-Rydberg atom
$(n-n')$-mixing processes have also been analyzed in \citet{bar07} and \citet{mas09}.

We found no non-LTE models similar to that of \citet{ver81} for the Sun
that would have provided all data necessary for calculations as in \citet{mih05}, in relation to the
atmospheres of DB white dwarfs and other similar helium-rich stellar atmospheres. However, since the
influence of strong electromagnetic emission is the same in helium-rich atmospheres as in
the Sun, the distribution shape of the helium excited-atom population probably deviates
considerably from the Boltzmann distribution. Taking this into account and because the data on the
relative efficiencies of atom-Rydberg atom processes and the relevant concurrent
processes would be useful for future non-LTE stellar-atmosphere models, the
chemi-ionization processes in $\textrm{He}^{*}(n)+\textrm{He}$ collisions were studied in
\citet{mih03} in the context of DB white dwarf atmospheres, within the
models proposed by \citet{koe80}. In the current study, for the same reason and using the same
models, the efficiencies of the excitation processes
%%%%%%%%%%%%%%%%%%%%%%%%%%%%%%%%%%%%%%%%%%%%%%%%%%%%%%%%%%%%%%
\begin{equation}
\label{eq:aa1}
\textrm{He}^{*}(n)+\textrm{He} = \left\lbrace
\begin{array}{ll}
\displaystyle{ \textrm{He}^{*}(n'=n + p)+\textrm{He},}\\
\displaystyle{ \qquad \qquad \qquad \qquad \qquad p\ge 1,}\\
\displaystyle{ \textrm{He}+\textrm{He}^{*}(n'=n + p),}
\end{array}
\right.
\end{equation}
%%%%%%%%%%%%%%%%%%%%%%%%%%%%%%%%%%%%%%%%%%%%%%%%%%%%%%%%%%%%%%
and the inverse deexcitation processes
%%%%%%%%%%%%%%%%%%%%%%%%%%%%%%%%%%%%%%%%%%%%%%%%%%%%%%%%%%%%%%
\begin{equation}
\label{eq:aa2}
\textrm{He}^{*}(n)+\textrm{He} = \left\lbrace
\begin{array}{ll}
\displaystyle{ \textrm{He}^{*}(n'=n - p)+\textrm{He},}\\
\displaystyle{ \qquad \qquad \qquad \qquad \qquad 0<p\le n-3,}\\
\displaystyle{ \textrm{He}+\textrm{He}^{*}(n'=n - p),}
\end{array}
\right.
\end{equation}
%%%%%%%%%%%%%%%%%%%%%%%%%%%%%%%%%%%%%%%%%%%%%%%%%%%%%%%%%%%%%%
(where $n+p$ and $n-p$ are the principal quantum numbers of the final Rydberg
states) are compared for $3\le n \le 8$ with the efficiencies of the relevant
concurrent processes. Here, we consider the well-known electron-excited-atom
collision excitation/deexcitation processes
%%%%%%%%%%%%%%%%%%%%%%%%%%%%%%%%%%%%%%%%%%%%%%%%%%%%%%%%%%%%%%
\begin{equation}
\label{eq:ea}
\textrm{He}^{*}(n)+\overrightarrow{e} =\textrm{He}^{*}(n \pm p)+\overrightarrow{e'},
\end{equation}
%%%%%%%%%%%%%%%%%%%%%%%%%%%%%%%%%%%%%%%%%%%%%%%%%%%%%%%%%%%%%%
where $\overrightarrow{e}$ and $\overrightarrow{e}'$ denote a free electron in
the initial and final states, respectively. The necessary calculations were performed
using the models in \citet{koe80} for DB white dwarf atmospheres with
$\log \, g = 8$ and $T_{eff}=12000$ K and $T_{eff}=14000$ K.

\section{Theoretical remarks}

As several of the above mentioned studies have reported, the same dipole resonant
mechanism (DRM) causes the non-elastic $(n-n')$-mixing and chemi-ionization processes
in $A^{*}(n)+A$ collisions. The DRM was also discussed in detail in \citet{mih12}. For
the $(n-n')$-mixing processes (1) and (2), with $A=\textrm{He}$, the DRM assumes that they
occur in the region $R_{n;n\pm p} \ll r_{n}$, where $R$ is the internuclear
distance and $r_{n}\sim n^{2}$ is the characteristic radius of atom $\textrm{He}^{*}(n)$,
in the vicinity of the corresponding resonant points $R_{n;n\pm p}$. These
parameters are found to be the roots of the equation $U_2(R)$ - $U_1(R) =
\varepsilon_{n\pm p} - \varepsilon_{n}$, where $\varepsilon_{n}$ and
$\varepsilon_{n\pm p}$ denote the energies of the Rydberg states $|n>$ and $|n\pm p>$
with the principal quantum numbers $n$ and $n \pm p$; and $U_{1}(R)$ and
$U_{2}(R)$ are the adiabatic energies of the electronic ground state $|1;R>$
and the first excited state $|2;R>$ of the molecular ion $\textrm{He}_{2}^{+}$. We took the
potential curves $U_{1}(R)$ and $U_{2}(R)$ from \citet{ign09} and following
the foundational work of \citet{jan79}, which focuses on the
$(n-n')$-mixing processes, we assumed that the processes (1) and (2)
are caused exclusively by the transitions $|n> \rightarrow |n+p>$ and $|n>
\rightarrow |n-p>$ of the outer electron, which occur simultaneously with the
transitions $|2;R>\rightarrow |1;R>$ and $|1;R> \rightarrow |2;R>$, respectively, in the
subsystem $\textrm{He}^{+} + \textrm{He}$, close to the resonant points $R_{n;n\pm p}$.
This fact is responsible for the resonant character of the processes in (\ref{eq:aa1}) and
(\ref{eq:aa2}).

For $A=$ H only chemi-ionization processes with $n\ge 4$ and
$(n-n')$-mixing processes with min$(n,n')\ge 4$ can be described on the basis of the DRM,
which caused by the existence of the stable negative ion $\textrm{H}^{-}$.
In contrast, for $A=$ He, the excited helium state with $n=3$
can be taken as the lower boundary of the block of Rydberg states. This was also used in \citet{mih03}, who focused
on the chemi-ionization processes in DB white dwarf atmospheres.

The processes (\ref{eq:aa1}) and (\ref{eq:aa2}) are characterized by the excitation and
deexcitation rate coefficients $K_{n;n+p}(T)$ and $K_{n;n-p}(T)$,
where $T$ is the local temperature of the atomic particles in the considered
stellar atmosphere. In accordance with the aim of this work, these rate
coefficients are defined as the quantities that determine the mean rates of transition
(per single excited atom) caused by the processes (\ref{eq:aa1}) and (\ref{eq:aa2}) between
whole shells with a given $n$ and $n\pm p$. They are determined here using a
method similar to that described in \citet{mih08}: the excitation rate coefficients $K_{n;n+p}(T)$ are
calculated directly and numerically in the semi-classical way, while the deexcitation
rate coefficients $K_{n;n-p}(T)$ are determined later according to the principle of
thermodynamical balance. All necessary expressions are given in \citet{mih08}.

For the DB white dwarf atmospheres considered here,
the values of $K_{n;n+p}(T)$ and $K_{n;n-p}(T)$, which are used in Eqs. (1) and
(2) for all necessary temperatures, can be calculated in principle by
fitting the values of $K_{n;n+p}(T)$ presented in Table 2 of \citet{mih08},
but only for the region $4\le n\le 8$. Bearing this in mind, Table
\ref{tab:rate} shows the values of the excitation rate coefficients
$K_{3;n+p}(T)$ when $1\le p \le 5$. These values, together with Table 1 in
\citet{mih08}, enable the calculation of $K_{n;n+p}(T)$ and $K_{n;n-p}(T)$ for the
entire block, $3\le n\le 8$.

The concurrent electron-Rydberg atom processes (3) are characterized here by
the rate coefficients $\alpha_{n;n\pm p}(T_{e}=T)$, where $T_{e}$ denotes the
electron temperature, which in the general case is not equal to the
atomic temperature. In the current study, the values of $\alpha_{n;n\pm p}(T)$ are
determined from the corresponding expressions given in \citet{vri80} and
\citet{joh72}. We note that ion-atom non elastic collisional processes, although
they are also characterized by long-range interaction, cannot be concurrent. The
reason is the huge difference in masses of electron and ion, which causes the
impact ion-atom velocity to be several orders of magnitudes lower than electron-atom
impact velocity.

\begin{figure}
\begin{center}
\includegraphics[width=\columnwidth,
height=0.95\columnwidth]{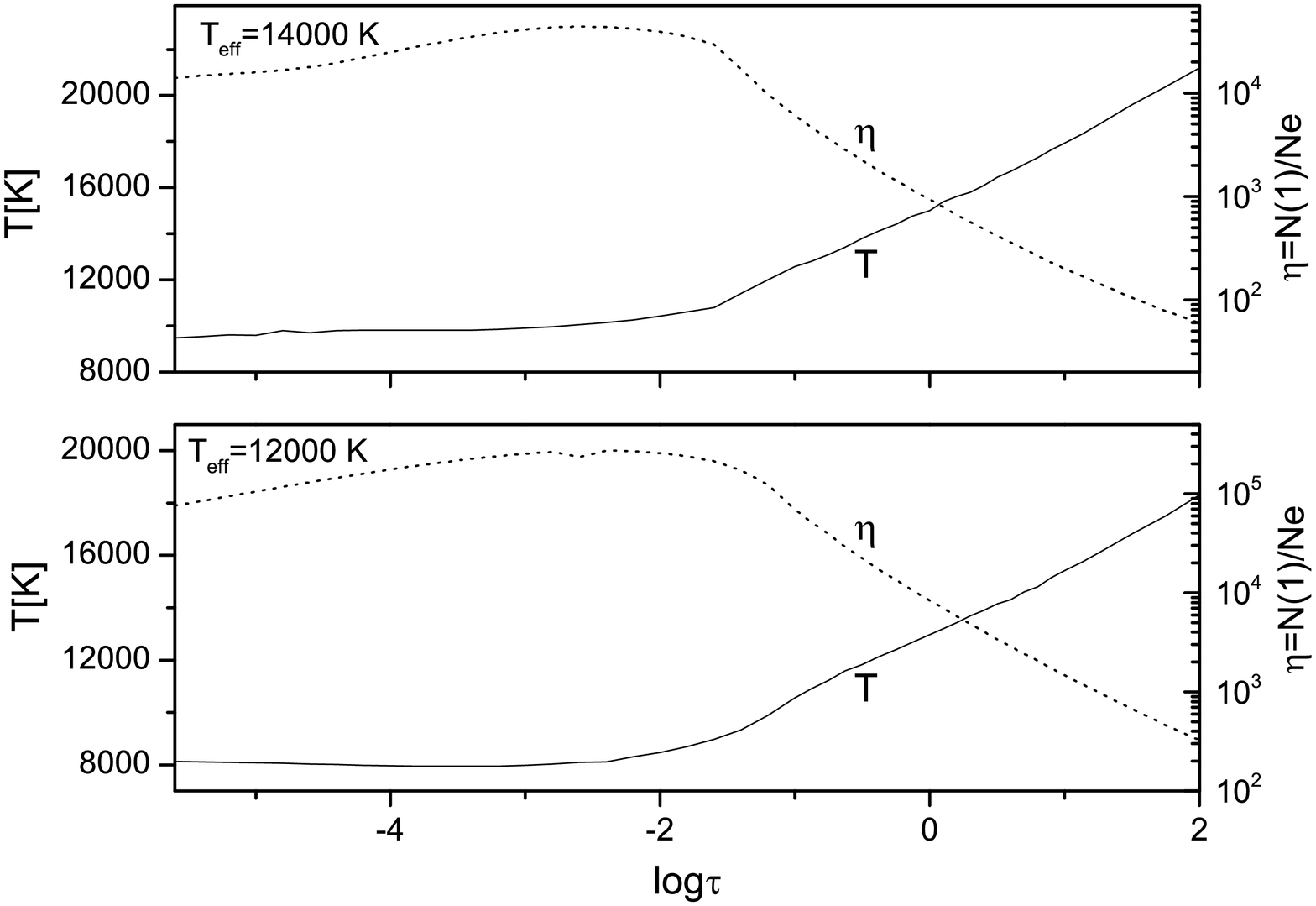}
\caption{Parameters (temperature $T$ - full line and
$\eta=N(1)/N_{e}$ - dotted) for DB
white dwarf atmosphere models with $\log \, g=8$ and $T_{eff}=12000$ K
and $14000$ K from \citet{koe80} as functions of $\log \, \tau$, where $\tau$ is
the Rosseland optical depth.}
\label{fig:DBpar}
\end{center}
\end{figure}

To estimate the relative efficiency of processes (\ref{eq:aa1}) and (\ref{eq:aa2})
and the concurrent processes (\ref{eq:ea}), it is sufficient to study the behavior of quantities
$F_{n}^{(+)}$ and $F_{n}^{(-)}$ in the considered DB white dwarf atmospheres, which are
given by the expressions
%%%%%%%%%%%%%%%%%%%%%%%%%%%%%%%%%%%%%%%%%%%%%%%%%%%%%%%%%%%%%%
\begin{equation}
F_{n}^{(+)}=\frac{\sum_{p=1}^{5} K_{n,n+p}\cdot N(n)\cdot N(1)}
{\sum_{p=1}^{5} \alpha_{n,n+p}\cdot N(n)\cdot N_{e}} =
\frac{\sum_{p=1}^{5} K_{n,n+p}}{\sum_{p=1}^{5} \alpha_{n,n+p}}\cdot\eta,
\quad n\ge 3,
\label{eq:F+}
\end{equation}
%%%%%%%%%%%%%%%%%%%%%%%%%%%%%%%%%%%%%%%%%%%%%%%%%%%%%%%%%%%%%%
\noindent and
%%%%%%%%%%%%%%%%%%%%%%%%%%%%%%%%%%%%%%%%%%%%%%%%%%%%%%%%%%%%%%
\begin{equation}
F_{n}^{(-)}=\frac{\sum_{p=1}^{n-3} K_{n,n-p}\cdot N(n)\cdot N(1)}
{\sum_{p=1}^{n-3} \alpha_{n,n-p}\cdot N(n)\cdot N_{e}} =
\frac{\sum_{p=1}^{n-3} K_{n,n-p}}{\sum_{p=1}^{n-3} \alpha_{n,n-p}}\cdot\eta,
\quad n\ge 4,
\label{eq:F-}
\end{equation}
%%%%%%%%%%%%%%%%%%%%%%%%%%%%%%%%%%%%%%%%%%%%%%%%%%%%%%%%%%%%%%
where $N(n)$, $N(1)$, and $N_{e}$ are the local densities of all
Rydberg helium atoms in the states with a given $n\ge3$, of a helium atom
in the ground state, and of free electrons, respectively; and the products
$K_{n,n \pm p}\cdot N(n)\cdot N(1)$ and $\alpha_{n,n \pm p}\cdot N(n)\cdot N_{e}$
are the partial atom- and electron-Rydberg atom excitation/deexcitation fluxes.
The factor $\eta$ is defined here by the relation
% %%%%%%%%%%%%%%%%%%%%%%%%%%%%%%%%%%%%%%%%%%%%%%%%%%%%%%%%%%%%%
\begin{equation}
\eta \equiv N(1)/N_{e} \cong (ionization \; degree)^{-1},
\label{eq:eta}
\end{equation}
%%%%%%%%%%%%%%%%%%%%%%%%%%%%%%%%%%%%%%%%%%%%%%%%%%%%%%%%%%%%%%
and its behavior in the atmospheres of the considered DB white dwarfs
(which is very important for the studied processes) is presented in Fig. \ref{fig:DBpar}.

\section{Results and discussion}

\begin{figure}
\begin{center}
\includegraphics[width=\columnwidth,
height=0.95\columnwidth]{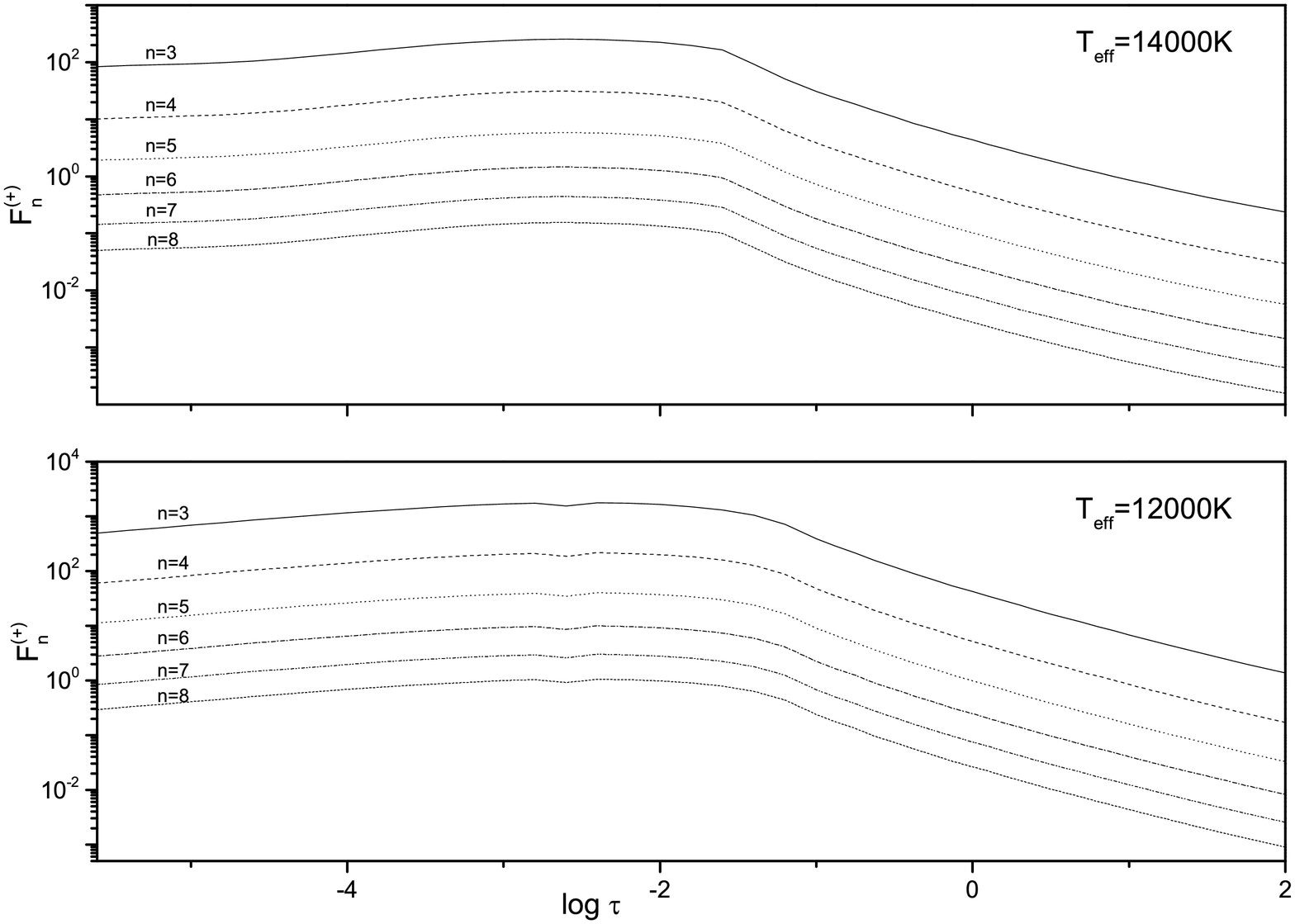}
\caption{Behavior of quantities $F_{n}^{(+)}$, given by
Eq.(\ref{eq:F+}) with $3\le n\le 8$, on the basis of DB white dwarf atmosphere
models from \citet{koe80} for $\log \, g=8$ and $T_{eff}=12000$ K and $14000$ K,
as functions of $\log \, \tau$, where $\tau$ is the Rosseland optical depth.}
\label{fig:DB1}
\end{center}
\end{figure}

Fig. 2 illustrates the behavior of the quantity $F_{n}^{(+)}$ with $3\le n\le 8$ in the
DB white dwarf atmospheres described by the models presented in \citet{koe80},
for $\log \, g = 8$ and $T_{eff}=12000$ K and $14000$ K. All results were obtained
using the data from \citet{koe80}, and these quantities are presented as functions of $\log \tau$, where $\tau$ is
the Rosseland optical depth. Fig. \ref{fig:DB1} unambiguously shows that for
DB white dwarf atmospheres with $T_{eff} \lesssim 14000$ K, the excitation
processes (\ref{eq:aa1}) in the lower part of the Rydberg block ($3\le n\le 8$) must be
taken into account as a new important factor that influences the helium
Rydberg-state populations. Fig. \ref{fig:DB1} shows that in the
parts of the atmosphere where the $\textrm{ionization degree is}\lesssim 10^{-3}$, the processes (\ref{eq:aa1})
with $n=3,4,$ and $5$ are dominant with respect to the known concurrent processes (\ref{eq:ea}); for
$n=6$ and $7$ they are dominant in the parts where the $\textrm{ionization degree is}\lesssim
10^{-4}$; and in the remaining parts their efficiency is comparable with
the efficiency of the processes (\ref{eq:ea}). Even the efficiency of the
processes (\ref{eq:aa1}) with $n=8$ is close to, or at least comparable with, the efficiency
of the processes (\ref{eq:ea}) in the parts where the $\textrm{ionization degree is} < 10^{-4}$.

These facts are especially important for the considered DB white dwarf atmospheres,
because for any temperature from the relevant temperature regions, the
corresponding Boltzman distribution of the excited helium atom populations will have
a distinct minimum (known as the "bottleneck") only in the region $3\le n <8$. Thus, it
is correct to expect that even a considerable perturbation of the mentioned distribution would not
result in a significant change in the position of this minimum. Hence, it follows that the
excitation processes (\ref{eq:aa1}) should by all means be taken into account in any future
non-LTE model of DB white dwarf atmospheres with similar parameters.

Examining the behavior of the quantities $F_{n}^{(-)}$ with
$4\le n\le 8$ in the same DB white dwarf atmospheres would lead to the same
conclusions, including the necessity of introducing the deexcitation processes (\ref{eq:aa2})
into the mentioned models. For this reason, the efficiencies of the considered $(n-n')$-mixing
processes are only illustrated in Fig. \ref{fig:DB1}.

Finally, we emphasize that these results are relevant not only to the
atmospheres of the DB white dwarfs mentioned here, but also to
a considerably wider range of helium-rich stellar atmospheres. For instance, an entire array
of white dwarfs of other types, as described in
\citet{weg85}, \citet{duf06}, and \citet{duf07}, have similar parameters. Consequently, processes
(\ref{eq:aa1}) and (\ref{eq:aa2}) must be just as important as the concurrent processes
(\ref{eq:ea}). Overall, the findings suggest that the considered atom-Rydberg atom
$(n-n')$-mixing processes (\ref{eq:aa1}) and (\ref{eq:aa2}) should be included \emph{ab initio}
in the modeling of the helium-rich white dwarf atmospheres.
%%%%%%%%%%%%%%%%%%%%%%%%%%%%%%%%%%%%%%%%%%%%%%%%%%%%%%%%%%%%%%%%%%%%%%%%%%%%%%%
%%%%%%%%%%%%%%%%%%%%%%%%%%%%%%%%%%%%%%%%%%%%%%%%%%%%%%%%%%%%%%%%%%%%%%%%%%%%%%%
%%%%%%%%%%%%%%%%%%%%%%%%%%%%%%%%%%%%%%%%%%%%%%%%%%%%%%%%%%%%%%%%%%%%%%%%%%%%%%%%%
\begin{table}[htbp]
\begin{center}
\caption{Excitation rate coefficients $K_{3;3+p}(T)$ for $1 \le p \le 5$ in $[10^{-9}\textrm{cm}^{3}\textrm{s}^{-1}]$} \label{tab:rate}
\begin{tabular}
{ c c c c c c c c } \hline
    \multicolumn{7}{c}{ $T$}\\
    & & & $[10^{3}$K] & & & & \\
   \cline{2-8}
  p&     5&      6&      8&     10&     12&     16&    20\\
  \hline
  1& 1.418&	1.723&	2.187&	2.514&	2.753&	3.074&	3.277\\
  2& 0.197&	0.271&	0.401&	0.504&	0.585&	0.701&	0.778\\
  3& 0.062&	0.091&	0.146&	0.192&	0.230&	0.286&	0.324\\
  4& 0.028&	0.042&	0.071&	0.096&	0.117&	0.148&	0.170\\
  5& 0.015&	0.023&	0.040&	0.055&	0.068&	0.088&	0.102\\
 \hline
 \end{tabular}
 \end{center}
\end{table}
%%%%%%%%%%%%%%%%%%%%%%%%%%%%%%%%%%%%%%%%%%%%%%%%%%%%%%%%%%%%%%%%%%%%%%%%%%%%%%%%%
\begin{acknowledgements}
The authors are grateful to the Ministry of Education, Science and
Technological Development of the Republic of Serbia for the support
of this work within the projects 176002, III4402.
\end{acknowledgements}

\bibliographystyle{aa}
%\bibliography{b}

\begin{thebibliography}{22}
\expandafter\ifx\csname natexlab\endcsname\relax\def\natexlab#1{#1}\fi

\bibitem[{{Barklem}(2007)}]{bar07}
{Barklem}, P.~S. 2007, \aap, 466, 327

\bibitem[{{Dufour} {et~al.}(2007){Dufour}, {Bergeron}, {Liebert}, {Harris},
  {Knapp}, {Anderson}, {Hall}, {Strauss}, {Collinge}, \& {Edwards}}]{duf07}
{Dufour}, P., {Bergeron}, P., {Liebert}, J., {et~al.} 2007, \apj, 663, 1291

\bibitem[{{Dufour} {et~al.}(2006){Dufour}, {Bergeron}, {Schmidt}, {Liebert},
  {Harris}, {Knapp}, {Anderson}, \& {Schneider}}]{duf06}
{Dufour}, P., {Bergeron}, P., {Schmidt}, G.~D., {et~al.} 2006, \apj, 651, 1112

\bibitem[{{Ignjatovi{\'c}} {et~al.}(2009){Ignjatovi{\'c}}, {Mihajlov}, {Sakan},
  {Dimitrijevi{\'c}}, \& {Metropoulos}}]{ign09}
{Ignjatovi{\'c}}, L.~M., {Mihajlov}, A.~A., {Sakan}, N.~M., {Dimitrijevi{\'c}},
  M.~S., \& {Metropoulos}, A. 2009, \mnras, 396, 2201

\bibitem[{{Janev} \& {Mihajlov}(1979)}]{jan79}
{Janev}, R.~K. \& {Mihajlov}, A.~A. 1979, \pra, 20, 1890

\bibitem[{{Janev} \& {Mihajlov}(1980)}]{jan80}
{Janev}, R.~K. \& {Mihajlov}, A.~A. 1980, \pra, 21, 819

\bibitem[{{Johnson}(1972)}]{joh72}
{Johnson}, L.~C. 1972, \apj, 174, 227

\bibitem[{{Koester}(1980)}]{koe80}
{Koester}, D. 1980, \aaps, 39, 401

\bibitem[{{Mashonkina}(2009)}]{mas09}
{Mashonkina}, L. 2009, Physica Scripta Volume T, 134, 014004

\bibitem[{{Mihajlov}(1982)}]{mih82}
{Mihajlov}, A.~A. 1982, in Proceedings of international conference on plasma
  physics. G 1982.

\bibitem[{{Mihajlov} {et~al.}(1996){Mihajlov}, {Dimitrijevi{\'c}}, \&
  {Djuri{\'c}}}]{mih96}
{Mihajlov}, A.~A., {Dimitrijevi{\'c}}, M.~S., \& {Djuri{\'c}}, Z. 1996,
  \physscr, 53, 159

\bibitem[{{Mihajlov} {et~al.}(2005){Mihajlov}, {Ignjatovi{\'c}}, \&
  {Dimitrijevi{\'c}}}]{mih05}
{Mihajlov}, A.~A., {Ignjatovi{\'c}}, L.~M., \& {Dimitrijevi{\'c}}, M.~S. 2005,
  \aap, 437, 1023

\bibitem[{{Mihajlov} {et~al.}(2004){Mihajlov}, {Ignjatovi{\'c}}, {Djuri{\'c}}, \&
  {Ljepojevi{\'c}}}]{mih04}
{Mihajlov}, A.~A., {Ignjatovi{\'c}}, L.~M., {Djuri{\'c}}, Z., \& {Ljepojevi{\'c}}, N.~N.
  2004, Journal of Physics B Atomic Molecular Physics, 37, 4493

\bibitem[{{Mihajlov} {et~al.}(2011){Mihajlov}, {Ignjatovi{\'c}},
  {Sre{\'c}kovi{\'c}}, \& {Dimitrijevi{\'c}}}]{mih11}
{Mihajlov}, A.~A., {Ignjatovi{\'c}}, L.~M., {Sre{\'c}kovi{\'c}}, V.~A., \&
  {Dimitrijevi{\'c}}, M.~S. 2011, \apjs, 193, 2

\bibitem[{{Mihajlov} {et~al.}(2008){Mihajlov}, {Ignjatovi{\'c}},
  {Sre{\'c}kovi{\'c}}, \& {Djuri{\'c}}}]{mih08}
{Mihajlov}, A.~A., {Ignjatovi{\'c}}, L.~M., {Sre{\'c}kovi{\'c}}, V.~A., \&
  {Djuri{\'c}}, Z. 2008, \jqsrt, 109, 853

\bibitem[{{Mihajlov} {et~al.}(1997){Mihajlov}, {Ignjatovi{\'c}}, {Vasilijevi{\'c}}, \&
  {Dimitrijevi{\'c}}}]{mih97}
{Mihajlov}, A.~A., {Ignjatovi{\'c}}, L.~M., {Vasilijevi{\'c}}, M.~M., \&
  {Dimitrijevi{\'c}}, M.~S. 1997, \aap, 324, 1206

\bibitem[{{Mihajlov} {et~al.}(2003){Mihajlov}, {Jevremovi{\'c}}, {Hauschildt},
  {Dimitrijevi{\'c}}, {Ignjatovi{\'c}}, \& {Alard}}]{mih03}
{Mihajlov}, A.~A., {Jevremovi{\'c}}, D., {Hauschildt}, P., {et~al.} 2003, \aap,
  403, 787

\bibitem[{{Mihajlov} {et~al.}(2007){Mihajlov}, {Jevremovi{\'c}}, {Hauschildt},
  {Dimitrijevi{\'c}}, {Ignjatovi{\'c}}, \& {Alard}}]{mih07}
{Mihajlov}, A.~A., {Jevremovi{\'c}}, D., {Hauschildt}, P., {et~al.} 2007, \aap,
  471, 671

\bibitem[{{Mihajlov} {et~al.}(2012){Mihajlov}, {Sre{\'c}kovi{\'c}}, {Ignjatovi{\'c}}, \&
  {Klyucharev}}]{mih12}
{Mihajlov}, A.~A., {Sre{\'c}kovi{\'c}}, V.~A., {Ignjatovi{\'c}}, L.~M., \& {Klyucharev},
  A.~N. 2012, Journal of Cluster Science, 23, 47

\bibitem[{{Vernazza} {et~al.}(1981){Vernazza}, {Avrett}, \& {Loeser}}]{ver81}
{Vernazza}, J.~E., {Avrett}, E.~H., \& {Loeser}, R. 1981, \apjs, 45, 635

\bibitem[{{Vriens} \& {Smeets}(1980)}]{vri80}
{Vriens}, L. \& {Smeets}, A.~H.~M. 1980, \pra, 22, 940

\bibitem[{{Wegner} \& {Koester}(1985)}]{weg85}
{Wegner}, G. \& {Koester}, D. 1985, \apj, 288, 746

\end{thebibliography}

\end{document}